# StructOpt: a modular materials structure optimization suite incorporating experimental data and simulated energies

Jason J. Maldonis[a], Zhongnan Xu[a], Zhewen Song[a], Min Yu[a], Tam Mayeshiba[a], Dane Morgan[a], Paul M. Voyles[a*]

[a] Department of Materials Science and Engineering, University of Wisconsin-Madison, 1509 University Ave., Madison, WI 53706, United States

**Abstract**

StructOpt, an open-source structure optimization suite, applies genetic algorithm and particle swarm methods to obtain atomic structures that minimize an objective function. The objective function typically consists of the energy and the error between simulated and experimental data, which is typically applied to determine structures that minimize energy to the extent possible while also being fully consistent with available experimental data. We present example use cases including the structure of a metastable Pt nanoparticle determined from energetic and scanning transmission electron microscopy data, and the structure of an amorphous-nanocrystal composite determined from energetic and fluctuation electron microscopy data. StructOpt is modular in its construction and therefore is naturally extensible to include new materials simulation modules or new optimization methods, either written by the user or existing in other code packages. It uses the Message Passing Interface's (MPI) dynamic process management functionality to allocate resources to computationally expensive codes on the fly, enabling StructOpt to take full advantage of the parallelization tools available in many scientific packages.

1. Introduction

The determination of the atomic structure of materials is a ubiquitous problem in materials science. Crystals with large unit cells, amorphous materials, and interfaces often have complex structure-property relationships that necessitate a thorough understanding of the material's atomic structure. In a typical approach to complex structure determination, a researcher would combine available data from experiments and simulations using scientific intuition, prior knowledge, and trial-and-error to discover the "best" structure for a given material. "Best" is often defined as the lowest energy structure most consistent with experimental data subject to (sometimes vague) constraints from prior knowledge. Unfortunately, available experimental data typically lack the information necessary to uniquely determine three-dimensional (3D) structure information, particularly for complex nanostructures. Pure simulation approaches, on the other hand, must often make approximations to reach realistic length and time scales at atomic resolution and therefore often cannot model realistic conditions giving rise to metastable structures or structures stabilized by complex environmental factors. Combining experimental data with simulations can provide the necessary spatio-temporal resolution and scale for systematic structure determination and allow structures to be determined that minimize energy as much as possible while remaining consistent with available experimental data.

In this context, the problem of materials structure determination can be redefined as a multi-objective, global optimization problem where the representation of a material (*e.g.* atomic structure and composition) is determined by an optimization algorithm that minimizes the system energy and the disagreement with experimental observations [1]. In general, an algorithm that simulates experimental data from an atomic structure (henceforth simply referred to as a *forward simulation*) yields simulated data and does not directly give the structure or structures that are consistent with a given data set, which is the "inverse modeling" problem. However, any forward simulation can be effectively turned into an inverse modeling tool by a sufficiently extensive search of the structure space to find all structures consistent with the available data. Unfortunately, there may be multiple structures that agree with the data, and the inversion can yield many structures that are unphysical in their energetics, e.g., have bond lengths too small/large or atoms in very unstable locations. By combining simulated energy information with forward simulations and experimental data, the atomic structures can be constrained more tightly, yielding ideally a unique or relatively small set of possible structures that are both energetically stable and consistent with experimental data.

Optimization methods have already been used to solve a variety of structure determination problems in science and engineering [2–6] including the prediction of globally stable [7–11] and metastable [12–14] crystal structures and amorphous materials [15–17]. Often the goal of these structure determination methods is to predict the lowest energy structure of a given stoichiometry under specific thermodynamic constraints [18,19]. In addition, there are a number of applications of optimization approaches for structure optimization incorporating forward simulations of experiments, e.g., in glass alloy structure determination [20–23]. The challenges of such optimization often requires the quantitative comparison to various sources of experimental characterization data [6], meaning "goodness-of-fit" values for multiple types of forward simulations need to be integrated into the optimization. Meredig and Wolverton have explored approaches very like those used here that integrate both simulated atomistic energy and forward simulation to define an objective function, *e.g.*, incorporating experimental X-ray diffraction data to guide a genetic algorithm to determine the lowest energy crystal structures of specific materials [24].

Here we present a flexible community tool to enable multiobjective structure optimization with guidance from both atomistic modeling-based energy prediction and multiple types of experimental forward simulations called *StructOpt* (short for Structure Optimizer). The primary goal and features of StructOpt are shown in Figure 1. StructOpt optimizes a population of atomic systems in search of structures that are energetically stable and show good agreement between forward simulations and experimental characterization data. The stability of the structure can be calculated at various levels of theory by atomistic modeling methods, although interatomic potentials of some kind are the most common approach [25]. Forward simulations are used to predict images or spectra of an atomic structure associated with a given characterization method and also can be implemented at various levels of fidelity [22,26–28]. StructOpt incorporates both types of calculations by constructing an objective function to evaluate an atomic structure's energy and agreement of multiple forward simulations to experimental data. An optimization algorithm is then built around this objective function to search the domain space for suitable solutions. StructOpt can evaluate multiple solutions simultaneously, allowing parallelized population-based searches for faster convergence and a more comprehensive search. The concept for StructOpt (and the name) was originally developed for genetic algorithm optimization of point defect clusters in crystals [29], then adapted to structure determination of single-specie nanoparticles against total energy and STEM data [30]. Those single-purpose research codes have been rewritten into the modular, extensible framework described here. We expect the StructOpt code to be useful for researchers attempting to determine complex 3D nanostructures from experimental data that cannot be directly inverted ((S)TEM, XRD, *etc.*).

The remainder of this paper is organized as follows: Section 2 gives an overview of StructOpt in the context of population-based optimization schemes such as genetic algorithms (GA). Section 3 demonstrates StructOpt's capabilities by providing example structure refinements against STEM and FEM data. Sections 4 and 5 describe the hierarchal and modular architecture of the code, which allows it to be easily extensible to new types of data and materials problems. Finally, Section 6 describes the parallelization techniques that allow StructOpt to scale to large numbers of cores and seamlessly incorporate parallelization in existing codes to calculate system energies or to simulate experiments.

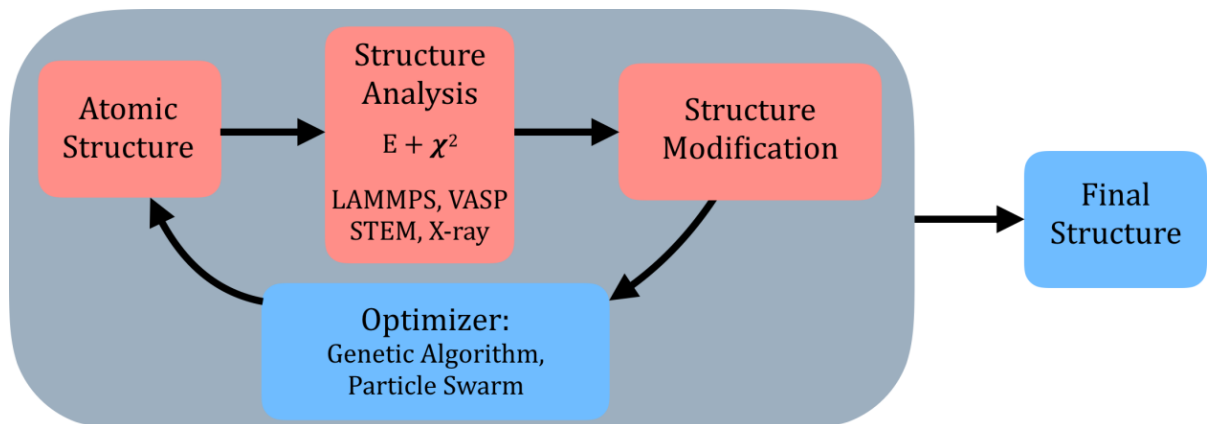

**Figure 1**: Schematic of atomic structure refinement program *StructOpt*. $E$ is the energy and $\chi^2$ represents the error in forward simulations compared to experiments.

## 2. Overview

Structure determination algorithms iteratively refine an atomic model by modifying the structure—for example by changing the positions of atoms—until the structure minimizes an objective function. StructOpt currently supports two structure determination algorithms [31]: a genetic algorithm [32] and a particle swarm optimization [33,34]. Both algorithms are population-based optimization schemes, allowing the features of the best candidate structures to collectively inform the creation of new candidates. These approaches allow for a more comprehensive search of the domain space given a sparse and diverse sampling of the initial population [32] than optimization approaches that operate on a single atomic structure. Genetic algorithms have been applied to atomic systems to study atomic clusters [35–38], crystal unit cells [9,10,39–41], defects [29], and grain boundaries [42,43]. The particle swarm optimization implemented in StructOpt was generally found to be slower and less adept at finding deep minima in energy-only optimizations than the genetic algorithm, so we focus on the genetic algorithm in this manuscript.

Given an atomic configuration $\theta$, an objective function is calculated via

$$O(\theta) = \sum_{i}^{N} \alpha_i f_i(\theta, D, C), \qquad (1)$$

where $f_i$ are specific terms of the objective function that evaluate a specific property of $\theta$. For any given application, at least one $f_i$ is likely to be an evaluation of energy. Additional $f_i$ can be defined that are related to experimental data ($D$) or user specified structural constraints ($C$). $\alpha_i$ are weighting factors chosen to give a desired numerical importance to each component of the objective function, typically set so that different contributions make similar levels of contributions to $O(\theta)$. Any number of terms can be included in the objective function (see Section 5). Alternative multi-objective optimization schemes exist that remove the need to explicitly set each $\alpha_i$ (such as Bayesian [44] and Pareto front [45] optimization), although these involve greater code and operation complexity and sometimes higher computation expense. Such tools are not presently available in StructOpt. $O(\theta)$ is minimized using the selected optimization algorithm to obtain atomic structure(s) with low values of each term in $O(\theta)$. In the context of genetic algorithms, the individual objective function terms are often called *fitness functions*, and their numerical output is called a *fitness score*. We use this naming convention here.

StructOpt currently supports optimization of atomic structures with periodic or non-periodic boundary conditions. Other constraints can be readily added due to the modular structure of the code. These constraints can help to define the type of structure that can be realized, such as nanoclusters with 3D surfaces, amorphous materials with or without periodic boundary conditions, and crystals with various symmetry constraints. All the operations within StructOpt can be applied to any type of structure, although some operations are more useful for one structure type than another. For example, an operation that moves surface atoms on a nanocluster may not have any surfaces on which to apply this move for a structure with periodic boundary conditions in all directions. The constraints on the structure can be defined by the user and can be as simple or complicated as the nature of the problem requires.

StructOpt defines a set of operations that can be applied to specific structures. In the context of genetic algorithms, these include crossovers, mutations, relaxations, and fitnesses. After creating or loading a set of atomic structures, the genetic algorithm performs crossovers on the population. Crossovers combine two structures by patching them together in unique ways, and result in two new structures that are added to the population. StructOpt then performs mutations on a subset of the population's structures, which modify a single structure to create a single new structure. Because crossovers and mutations can create unstable structures, relaxations are performed to move the atomic configuration towards a local minimum of one or more parts of the objective function (for example, some relaxations are designed to bring the modified structures into energetic equilibrium). Fitness evaluations require forward simulations performed on a static structure and provide information about the "goodness" of the structure. Fitness evaluations are run after the relaxation process. The weighted sum of the fitness evaluations in Eq 1 provides a single fitness score for each individual. Once all structures have been assigned a fitness score, StructOpt uses this information to remove certain structures from its current population and repeats the above processes. This iterative procedure is repeated until the convergence criteria—usually based on the fitness scores—are satisfied. StructOpt does not limit these operations (crossovers, mutations, relaxations, forward simulations, and structure selectors) to genetic algorithms, so they can be used within other optimization algorithms defined in StructOpt or defined by the user.

Crossovers and mutations modify the atomic structure to sample both nearby and distant parts of the configuration space. Given a diverse starting population, crossovers allow for sampling new areas of the search space by combining two distinct solutions, whereas mutations are perturbations on an existing structure. The crossovers implemented in StructOpt include a crossover that rotates two models, cuts both rotated models along a plane, and glues the opposing slices together, producing two children [35]. The mutations in StructOpt are numerous and include swapping atom species, randomly moving a single atom, randomly moving a cluster of atoms, moving atoms from the bulk to the surface, and twisting the model. Mutations and crossovers are among the simplest operations to add to StructOpt, and they can be configured to work differently on different structure types.

StructOpt implements three structural relaxation approaches, which are primarily used to relax locally unstable atomic configurations that result from crossovers, mutations, *etc*. The first is to relax energy using the molecular simulation package LAMMPS [5]. StructOpt can use any minimization algorithm provided by LAMMPS to relax the atomic structure, including conjugate gradient minimization or molecular dynamics. The second relaxation mechanism is a hard-sphere relaxation that simply moves atoms to ensure that no two atoms overlap a fixed radius around each atom. Finally, StructOpt has implemented a special relaxation of atomic positions related to STEM image matching, which we call the "STEM relaxation". The STEM relaxation performs a rotation of the atomic model to best match the STEM image without changing the positions of the atoms relative to one another.

StructOpt currently implements three objective function terms. The first term, called LAMMPS inside StructOpt, finds the potential energy of the relaxed atomic arrangement using the LAMMPS package which requires the user to choose an appropriate interatomic potential. The optimization of the LAMMPS objective function term attempts to find the lowest energy arrangement, and if it is the only term in the objective function it will ideally yield the zero-temperature ground state of the applied interatomic potential for the given constraints. The second routine, called STEMSIM, provides an algorithm for comparing atomic structures to STEM images. STEMSIM [26] calculates the discrepancy in pixel intensities between a STEM experimental image and a simple linear convolution STEM image simulation model [38,39]. If STEMSIM is the only term in the objective function, then the ideal optimized structure will have atom positions consistent with the experimental STEM image along one 2D orientation, but the bond lengths in other orientations and atomic forces and energies will be unconstrained and may be unrealistic. FEMSIM calculates the discrepancy between fluctuation electron microscopy (FEM) experimental data and a FEM simulation using a kinematic diffraction approach [28,47] and is useful for amorphous systems. Similar to STEMSIM, if FEMSIM is the only term in the objective function then the optimized structure will be consistent with the experimental data but may have unphysical atom positions which are very far from an energy minimum [48]. Examples of the use cases of these objective terms can be found in the examples presented in Section 3.

After the objective functions of all individuals in the population are calculated, the final task of the genetic algorithm is deciding which individuals are discarded in the next generation. StructOpt includes selection and predator operations for choosing which individuals to mate and keep, respectively, by evaluating their objective function. StructOpt comes with several operators that affect the diversity of the population. Selection and predator operators such as "best", "rank", and "roulette" promote the dissemination and retention of the best individual's features, while "fuss" and "tournament" have the option of maintaining a more diverse population [49–51]. Predators can also be used to eliminate duplicate structures, thereby increasing diversity and eliminating premature convergence on a non-ideal structure.

The operations described above all require varying inputs, outputs, and resources. In the next sections, we discuss how StructOpt's object-oriented approach facilitates flexibility and extensibility of the entire codebase, the wrappers around StructOpt to make that it user-friendly, and the dynamic resource allocation system that handles the various parallelization requirements. We first present two examples that demonstrate how experimental STEM and FEM data can be used to refine the atomic structure to metastable particles and amorphous solids.

### 3. Examples

Figures 2 compares the optimization of just the energy term with the optimization of both the energy and STEMSIM terms for refining the structure of nanoparticles. Figure 3 shows a similar comparison using the FEMSIM term and the structure of amorphous materials. In both cases, simulated data is substituted for experimental data for better model assessment. Such simulated data is often called *phantom data* because the structure that is used to create the data is known *a priori*. Therefore, the target of the optimization is also known *a priori*. When the simulation finishes, the output structure(s) can be compared to the target structure to quantitatively determine the accuracy of the solution(s).

In Figure 2(a), StructOpt optimized the structure of a 561 atom Pt nanoparticle ($Pt_{561}$) using an embedded-atom method (EAM) force field potential [52]. The initial population consisted of 18 randomly generated fcc particles already in the (100) orientation, and the STEMSIM relaxation procedure was used to align the modified atomic structures to the STEM image throughout the optimization. When the optimization was constrained by only the system energy, StructOpt found the icosahedral $Pt_{561}$ particle, which is the lowest energy, magic number particle with 561 atoms [53,54]. The icosahedral structure is fcc-like with tilt boundaries that allow the surface to consist of low energy, hexagonal close-packed planes with a low surface-to-volume ratio. The final structure has minor surface defects with respect to the perfect icosahedral structure, but the overall icosahedral symmetry is found.

When a STEM fitness function is added to the objective function, StructOpt identifies a different, metastable optimal structure. In Figure 2(b), the objective function is modified to contain both an energetic term ($E_f$) and a term representing the fit ($\chi^2$) to a simulated STEM image of a 561 atom cuboctahedron. The cuboctahedron shape has an fcc structure with no defects and exposes higher energy (100) surfaces captured in the phantom STEM image which do not exist in the icosahedron in Figure 2(a). When the additional constraint of error vs. the phantom is added to the objective function (with $\alpha_{LAMMPS} = 1$ and $\alpha_{STEMSIM} = 100$), the $Pt_{561}$ icosahedron is not found as the optimal structure and instead StructOpt identifies the cuboctahedron as the correct target structure. Note that the experimental image is simulated, so the STEM fitness at the end of the algorithm has a fitness value close to zero upon finding the optimal structure. The fitness value is not exactly zero because the identified structure has minor surface defects with respect to the target structure, but previous results using this technique show that for phantom data the atom positions are recovered nearly exactly [30]. The optimization using STEM data only has not been carried out, since the STEM data will not constrain the structure in the direction perpendicular to the plane of the image. As a result, the optimized structure is guaranteed to be unphysical [48].

Figure 3 shows results for the optimization of 256 atoms of Al with periodic boundary conditions using an Al EAM potential [55]. Similar to Figure 2(a), the optimization shown in Figure 3(a) is performed with just an energy term. In this case the fcc Al ground state is obtained, as expected. The optimization in Figure 3(b) includes both energy and forward simulation objective function terms, where the simulation is compared to phantom FEM data calculated from an amorphous/nanocrystal composite structure proposed as a model for high Al-content metallic glasses [56] (see [57], Fig. 1 for example experimental data on $Al_{88}Y_7Fe_5$). Since the target structure is a model of a glass, it is metastable in energy. The initial population for the genetic algorithm contained purely amorphous structures, and the crossovers and mutations modified the structures in the population until an amorphous-nanocrystal composite was formed that matched the experimental data well and had low energy. The FEM data captures the nanometer-scale order information of the phantom composite material, driving the optimal structure away from the energetically optimal fcc Al single crystal (Figure 3(a)) and towards a metastable material that has better agreement with the FEM data (Figure 3(b)). The fraction of Voronoi indices that are crystal-like [23][ is similar in the final optimized structure and the phantom structure.

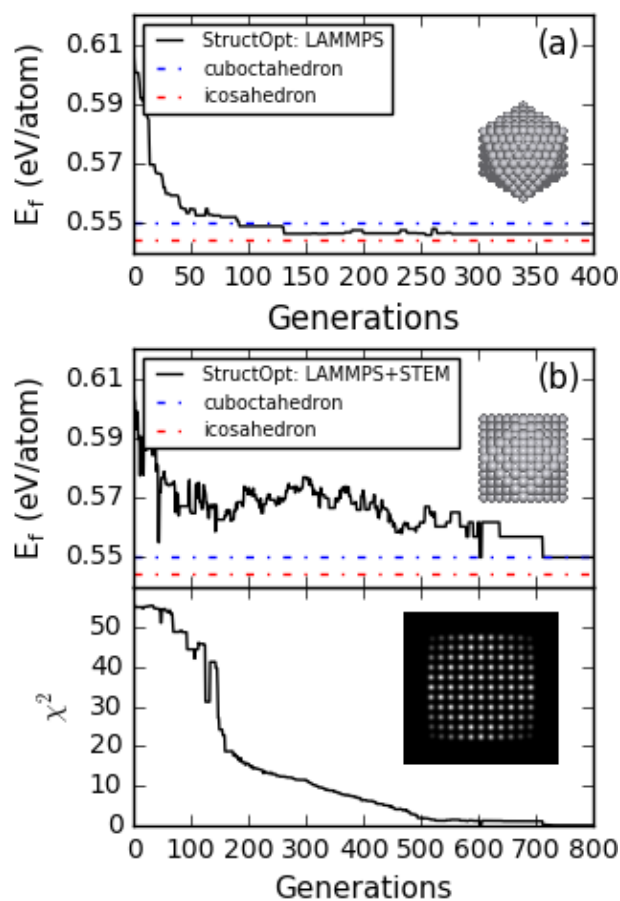

**Figure 2**: A StructOpt run optimizing the structure of a $Pt_{561}$ nanoparticle using an EAM potential (a) without and (b) with the STEMSIM term in the objective function. The inset in (a) shows the $Pt_{561}$ regular icosahedron with 20-fold symmetry found by StructOpt using only an energy term in the objective function. The inset in (b)-upper shows the $Pt_{561}$ cuboctahedron refined by StructOpt using both energy and STEMSIM terms in the objective function. Minimization of the relative stability, $E_f$, drives the energy of the system down. Meanwhile, the STEMSIM term ($\chi^2$) gives the magnitude of the error between the forward STEMSIM simulation and the STEM image in inset (b)-lower, and drives the system toward agreement with the phantom experimental data. The blue and red dashed lines show the relative formation energies of the $Pt_{561}$ icosahedron and cuboctahedron particles, respectively, calculated using an EAM potential and LAMMPS.

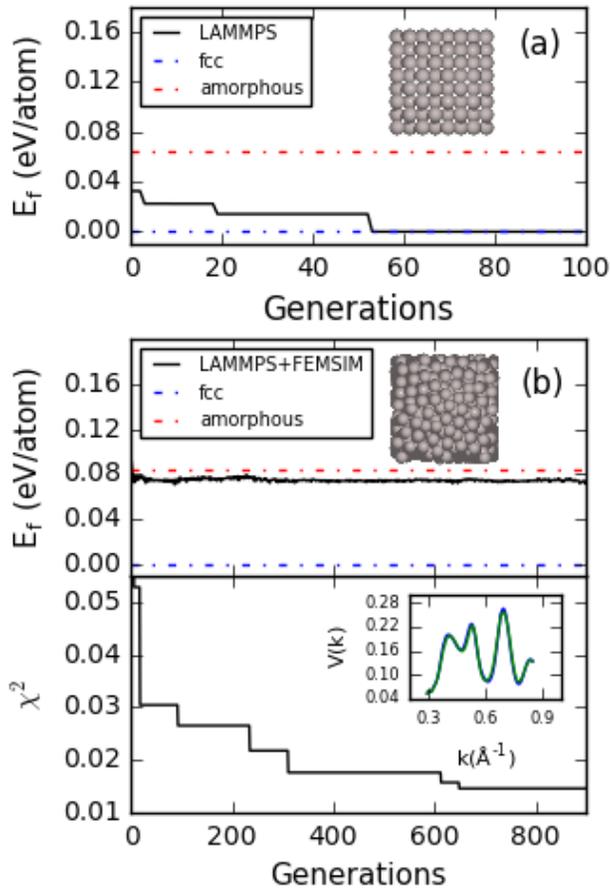

**Figure 3**: A StructOpt run optimizing the structure of Al using an EAM potential (a) without and (b) with the FEMSIM fitness function and phantom data from a nanocrystal / amorphous composite. The black lines show the fitness values calculated by StructOpt using LAMMPS and FEMSIM. The blue and red dashed lines show the formation energies of an Al fcc crystal and the nanocrystal / amorphous composite target, respectively, calculated using an EAM potential in LAMMPS. The inset in (a) shows the $Al_{256}$ fcc structure found with an energy-only optimization. The inset in (b)-upper shows an $Al_{3692}$ amorphous structure refined by StructOpt when the FEM data in inset (b)-lower is incorporated into the FEMSIM cost function. In the inset in (b)-lower, the blue line is the FEMSIM data of the phantom target and the green line is the FEMSIM data of the optimal structure StructOpt identified.

4. Code Architecture

StructOpt is written in python and is open-source. It was designed with modularity and flexibility in mind, and its object-oriented approach makes clear distinctions between different types of functionality. The code has three important pieces of functionality, which are the *Individual* class, the *Population* class, and the *optimization* algorithm. The *Individual* class is a stand-alone object containing an atomic structure with the necessary functionality to operate on itself. The *Population* class contains a list of *Individual* instances and provides functionality for operating on *Individual* instances. The *optimization* algorithm uses the functionality of the *Population* class to iteratively modify the underlying individuals. It is the job of the optimizer to shape the individuals in the population into a form that satisfies the convergence criteria. By making clear distinctions between these different pieces of functionality, each part of the code can be used and modified independently.

The *Individual* class extends the *ASE.Atoms* [58] class and implements methods to modify an instance of itself in the context of the implemented optimization algorithms. Specifically, an *Individual* can mutate, relax, and perform various fitness evaluations on itself in the context of the genetic algorithm and can perform analogous operations for particle swarm. StructOpt makes use of *ASE*'s LAMMPS calculator to calculate energies from LAMMPS and to encourage code reuse. Because different types of atomic structures exist, StructOpt implements different classes that inherit from the *Individual* class. Subclasses of *Individual* implement methods that are specific to the atomic structure they are modeling, overriding any necessary functionality. The two types of structures currently implemented in StructOpt are the *Periodic* and *APeriodic* classes which enforce and do not enforce periodic boundary conditions, respectively. Materials problems that require nanoclusters can use the *APeriodic* structure type, while materials problems that require optimizing unit-cells of crystals or supercells of more complex materials can use the *Periodic* structure type. Figure 3 is an example of using a *Periodic* structure to optimize an amorphous-nanocrystal composite. Implementing new *Individual* classes with different boundary conditions or other constraints only requires inheriting from the *Individual* and making any necessary physically-motivated modifications to the code. For example, a new *Individual* class might be called *Surface*, which would have periodic boundary conditions in two dimensions. Once defined, the new *Individual* subclasses can be used immediately in any implemented optimization algorithm.

The *Population* class implements methods for operating on multiple *Individual* instances at once. For example, the *Population* can perform crossovers on and removals of the individuals in the population as required by the genetic algorithm. The *Population* also manages parallelized execution of the *Individual* methods. The complicating factor of this class is communication during parallelization because different parallelized processes can modify a *Population* instance in different ways. A more detailed description of the issues and implementation decisions around parallelization can be found in Section 6 and in the online documentation.

The goal of the optimization algorithm is to use the methods in the *Population* and *Individual* classes to modify the atomic structures into a form that satisfies the convergence criteria. Standard versions of genetic and particle swarm algorithms are included as examples, and users can modify these codes to apply additional optimization algorithms if desired.

StructOpt requires an input file specifying the details of the initial population, objective function terms, relaxation algorithms, output data, and parameters associated with the optimization algorithm, such as mutations, crossovers, and selection schemes for the genetic algorithm. These inputs parameterize the optimization and allow StructOpt to be highly configurable. This flexibility comes at the cost of increased complexity, so a hierarchical file format (*json*) was chosen to help mitigate some of the complexity. The output data of StructOpt simulations includes structure files, files for input/output of external programs such as LAMMPS, and logs of fitness scores, genealogy, and which crossovers/mutations were applied to specific atomic structures.

The output data can be parsed using a data exploration utility that is packaged with StructOpt called *DataExplorer*. In addition to providing standard data processing for reading outputs such as fitness scores, the data explorer provides utility functions for working with the genealogy of the population after genetic algorithm optimization and provides access to all the functionality within StructOpt (*e.g.* relaxations, crossovers, and fitness calculations). Any operation that can be run on an *Individual* instance within StructOpt can also be applied to an atomic model loaded through the data explorer. It is therefore straightforward to reevaluate the fitness of a particular structure, for example, to identify a crucial turning point in the simulation. Additionally, since the data explorer is modularized in much the same way as the core components of StructOpt, its analysis techniques can be tied directly into StructOpt and used to tune the parameters of an optimization on-the-fly.

Finally, StructOpt also comes packaged with a simple job manager utility. This can be replaced by other workflow managers such as FireWorks [55], Pegasus [59], or CONDOR [60] but the built-in job manager provides a simple way to design a series of simulations and manage their results within a uniform environment. By combining the job manager with the data explorer, a single script can submit, track, and analyze a set of jobs for determining the behavior of the optimizer against various parameters. These capabilities are particularly useful for hyperparameter optimization, for example to test a variety of $\alpha$'s in Eq 1 to determine their effect on the optimized structures.

## 5. Extending the Code

There are three areas in which the codebase is designed for easy modification and extension. New optimization algorithms can be implemented that use the current functionality in new ways; additional constraints can be implemented to cover a wider range of structure types; and new functionality for the optimizations (such as crossovers, fitnesses, *etc*.) can be added. Each of these areas of modifications can be done independently. For example, to add the particle swarm optimization to the existing code implementing genetic algorithms, we added a new module that updates the atomic positions via the approaches from the particle swarm method [61], which replaced the genetic algorithm's crossovers and mutations. The other functionality of StructOpt was used without modification. Smaller-scale modifications to the optimization scripts are also designed to be fast and intuitive to the user. For example, in a genetic algorithm optimization, the number of individuals in a population can be controlled by updating a parameter to the *Population.kill* method over the course of the simulation.

As an illustration of how to implement a new optimization constraint, consider the example of a crystal constrained to a specific space group. Atoms could be restricted to an asymmetric unit within the unit cell, then a set of symmetry operations could be implemented to extend those atoms to fill the unit cell or a large supercell if necessary. New mutations and crossovers could be defined to operate only on well-defined atom sites rather than general atom positions, and a short wrapper that populates the sites with atoms could be wrapped around the pre-existing *Individual* fitness and relaxation methods to easily incorporate their pre-existing functionality. Once crossovers, mutations, and wrapper functions have been written, the *Population* class and optimization algorithms can use this new constraint without modification to their code.

Finally, crossovers, mutations, and other methods that are designed to modify atomic structures in unique ways can be added easily. Each method is contained within its own file and is attached to the *Individual* class (or a subclass thereof) via an import in the corresponding file. Then, the new operations are enabled within the input file by specifying the name of the operation.

While StructOpt is written in Python, its functionality is not limited to Python. External programs that operate on atomic structures can be incorporated into StructOpt. In particular, the fitness modules in StructOpt are all implemented in different languages: STEMSIM is implemented in native Python; LAMMPS is written in C++; and FEMSIM is written in FORTRAN. Any relaxation code is expected to accept an atomic structure as input and output a modified atomic structure. Similarly, any fitness operation is expected to accept an atomic structure and output a fitness score.

## 6. Parallelization and Resource Allocation

In addition to being easily extensible, the code architecture allows for flexible resource management. Depending on the computational expense and complexity of the algorithm, the user can instruct StructOpt to allocate one or more cores to the methods that operate on the individuals in the population. For example, a force-field calculation of a system's energy may require only one core, but a first-principles calculation may require many. The different computation algorithms and varying number of structures that need to be evaluated requires StructOpt to be capable of dynamically allocating its available resources at every iteration.

StructOpt utilizes two types of parallelization to handle resource allocation, shown schematically in Figure 6. The single-core-per-individual (SCPI) parallelization mechanism uses *mpi4py* [62–64] to distribute *Individual* instances over the available cores. If there are at least as many cores as *Individual* instances, each instance is assigned to a different core. If there are more *Individual* instances than cores, the instances will be divided amongst the available cores, and the instances that are assigned to the same core will be evaluated serially on that core. This method does not support parallelization to more cores than there are *Individual* instances. *mpi4py* handles the communication required to collect the updated information from each core and redistribute the updated *Individual* instances.

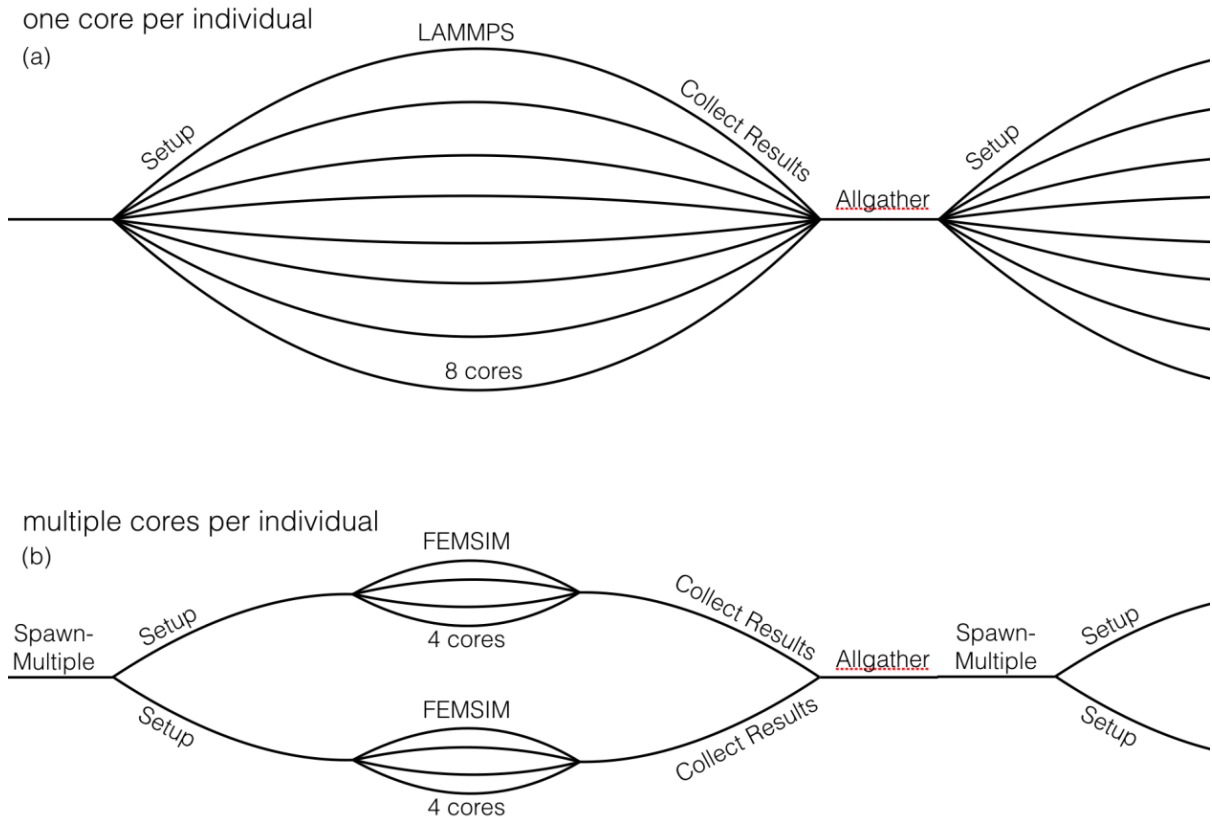

Figure 6: The SCPI (a) and MPMD (b) methods differ in the way in which they parallelize the code. Standalone operations that are computationally expensive benefit from MPI's MPMD functionality, while less computationally intensive operations can be parallelized at the population level using SCPI.

If an evaluation algorithm (such as FEMSIM) already takes advantage of MPI, it can be used within StructOpt via the multiple-program-multiple-data (MPMD) functionality supported by OpenMPI [65]. Under the MPMD settings, StructOpt uses the *MPI_Spawn_Multiple* command to allocate multiple cores to each *Individual* instance. When using MPMD, the user can specify a range of cores to be used for each evaluation. StructOpt attempts to intelligently assign a number of cores to each *Individual* instance to minimize the total computation time. For example, if the user specifies that an evaluation should use between 2 and 16 cores, and for a given iteration four individuals need to be evaluated, StructOpt will allocate four cores to each Individual; if, for another iteration, only one individual needs to be evaluated, StructOpt will allocate all 16 cores to that one individual.

On the other hand, some operations are fast enough that they should only be run on the master core. These operations are defined by StructOpt's *@root* Python decorator, which provides simple syntax for broadcasting the result of the operation on the master core to all other cores. For operations that are parallelized, the *allgather* operation in StructOpt collects the updated *Individuals* from each core, combines them into a single updated *Population*, and propagates the updated *Population* to all cores.

StructOpt's parallelization mechanisms allow a variety of different algorithms to be incorporated into its framework without significantly reducing each algorithm's parallelized potential. It is likely that the algorithms that will most benefit from parallelization will be relaxation or fitness algorithms that either intelligently move atoms or perform simulations of experiments. Users can integrate their pre-existing techniques and codes into StructOpt to solve their own unique structure determination problems.

Population-based optimization algorithms have the advantage that they can be highly parallelized. StructOpt takes advantage of this potential by combining both SCPI and MPMD parallelization schemes into a dynamic resource management system. SCPI and MPMD scale differently with the number of cores used. SCPI parallelizes up to a number of cores equal to the number of individuals in the population, while MPMD is limited by the scaling potential of the operation that is being parallelized (per individual). For example, assuming an MPMD program is efficient up to 100 cores and StructOpt uses a population size of 20, the code has the potential to scale up to 20*100=2,000 cores. The limitation to scaling of this type is the time spent outside the parallelized MPMD functions, which will reduce the speedup of the overall code at a high number of cores, following Amdahl's law. In addition, a combination of SCPI and MPMD processes may result in many cores being unused for a non-trivial amount of wall-clock time, in which case the scaling of StructOpt will be poor. In practice, we find that StructOpt parallelizes with a speedup of 50-100% up to one core per individual for a range of problems. The speedup drops significantly at higher numbers of cores, but in a way that is strongly problem-specific.

## 7. Conclusion

StructOpt, an open-source structure determination program, provides a framework for incorporating simulated energies and experimental characterization data into atomic structure optimization of materials. StructOpt is highly modular and can therefore easily integrate new types of experimental data, optimization algorithms, and optimization constraints into its infrastructure. StructOpt has already been used to solve complex nanoparticle and amorphous structure problems. It currently implements genetic algorithm and particle swarm optimizations, and can use LAMMPS, STEM, and FEM data within its objective function. In addition, StructOpt implements an advanced dynamic resource allocation system to take advantage of MPI functionality and existing parallelized applications for materials simulations.

**Acknowledgements**

Primary development and testing of StructOpt was supported by the University of Wisconsin-Madison College of Engineering Research Innovation Fund and the National Science Foundation DMREF program (DMR-1332851). Initial development of the StructOpt concept and single-use research codes was supported by a NSF Software Infrastructure for Sustained Innovation, DMR-1148011. Documentation, clean-up, and release of the code, and preparation of this manuscript was support by the NSF DMREF (DMR-1728933).

**Data Availability**

The raw data required to reproduce these findings are available to download from https://github.com/uw-cmg/StructOpt. The processed data required to reproduce these findings are available to download from https://github.com/uw-cmg/StructOpt.